\documentclass[showpacs,twocolumn,pra,reprint]{revtex4-1}

\usepackage{slashed}
\usepackage{mathrsfs}
\usepackage{amsmath}
\usepackage{amssymb}
\usepackage{revsymb}
\usepackage{graphicx,accents}
\usepackage{graphicx,epstopdf}
\usepackage{mathrsfs}
\usepackage{bm}
\usepackage{psfrag}
\usepackage{color}
\usepackage{hyperref}
\hypersetup{colorlinks=true,citecolor=Red,urlcolor=Red}
\usepackage{bbm}
\usepackage{bbold}
\usepackage{tabularx}
\usepackage{graphicx}
\usepackage{ulem}
\usepackage[usenames,dvipsnames]{xcolor}

\newcommand{\be}{\begin{equation}}
\newcommand{\ee}{\end{equation}}
\newcommand{\ba}{\begin{array}}
\newcommand{\ea}{\end{array}}
\newcommand{\bea}{\begin{eqnarray}}
\newcommand{\eea}{\end{eqnarray}}
\newcommand{\bd}{\begin{displaymath}}
\newcommand{\ed}{\end{displaymath}}

\newcommand{\tbf}[1]{\textbf{#1}}

\newcommand{\figref}[1]{Fig.~\ref{#1}} 
\newcommand{\eqnref}[1]{Eq.~(\ref{#1})}

\newcommand{\tsf}[1]{\textsf{#1}}

\newcommand{\vphi}{\varphi}

\newcommand{\e}{\mathbb{e}}

\makeatletter
\DeclareRobustCommand{\cev}[1]{%
  \mathpalette\do@cev{#1}%
}
\newcommand{\do@cev}[2]{%
  \fix@cev{#1}{+}%
  \reflectbox{$\m@th#1\vec{\reflectbox{$\fix@cev{#1}{-}\m@th#1#2\fix@cev{#1}{+}$}}$}%
  \fix@cev{#1}{-}%
}
\newcommand{\fix@cev}[2]{%
  \ifx#1\displaystyle
    \mkern#23mu
  \else
    \ifx#1\textstyle
      \mkern#23mu
    \else
      \ifx#1\scriptstyle
        \mkern#22mu
      \else
        \mkern#22mu
      \fi
    \fi
  \fi
}

\newcommand*\xbar[1]{%
  \hbox{%
    \vbox{%
      \hrule height 0.5pt 
      \kern0.2ex
      \hbox{%
        \kern-0.15em
        \ensuremath{#1}%
        \kern-0.15em
      }%
    }%
  }%
}

\newcommand{\BK}[1]{\textcolor{blue}{\tbf{#1}}}

\graphicspath{ {./figures/} }

\begin{document}
\title{One-photon pair-annihilation in pulsed plane-wave backgrounds}
\author{S.~Tang}
\email{suo.tang@plymouth.ac.uk}
\affiliation{Centre for Mathematical Sciences, University of Plymouth, Plymouth, PL4 8AA, United
Kingdom}

\author{A.~Ilderton}
\affiliation{Centre for Mathematical Sciences, University of Plymouth, Plymouth, PL4 8AA, United Kingdom}

\author{B.~King}
\affiliation{Centre for Mathematical Sciences, University of Plymouth, Plymouth, PL4 8AA, United Kingdom}


\begin{abstract}
We study the $2\rightarrow1$ process of electron-positron pair annihilation to a single photon in a plane-wave background.
The probability of the process in a pulsed plane wave is presented, and a locally constant field approximation is derived and benchmarked against exact results. The stricter kinematics of annihilation (compared to the $1\rightarrow2$ processes usually studied) leads to a stronger dependence on the incoming particle states. We demonstrate this by studying the effect that initial state wavepackets have on the annihilation probability. The effect of annihilation in a distribution of particles is studied by incorporating the process into Monte Carlo simulations.
\end{abstract}
\maketitle
%
%
%
\section{Introduction}
%
At the single-vertex level, the stimulated QED processes that can occur in a background electromagnetic field may be divided into the $1\rightarrow2$ processes, namely nonlinear Compton scattering (NLC)~\cite{nikishov64, kibble64} and nonlinear Breit-Wheeler (NBW)~\cite{nikishov64,berestetskii1982quantum} and the $2\rightarrow1$ processes of one-photon absorption~\cite{ritus85} and one-photon pair-annihilation~\cite{ritus85}. All these processes are forbidden in the absence of a background.

NLC and NBW have been thoroughly studied both analytically~\cite{nikishov64,kibble64,ritus85,Boca:2009zz,harvey09, PRA2011Seipt, PLB2012nousc, mackenroth11, king12b, PRA2016Harvey, king16b, dinu16, meuren17} and through numerical implementation in particle-in-cell (PIC) codes~\cite{PRE2015Gonoskov,Arber_2015}, where they contribute to e.g.~electromagnetic cascade formation~\cite{PRL2008Bell,PRL2010Fedotov,PRSTAB2011Elkina,PRL2012Ridgers,PRA2013king,TangPRA2014,PRA2019Ilderton}.  In this context the $2\rightarrow1$ processes are entirely neglected, the usual justification being that the outgoing particle phase space of these processes is completely determined by that of the initial particles, leading to their probabilities being proportional to an initial particle density factor that suppresses the process, relative to NLC and NBW. (The theory of one-photon pair-annihilation has been studied in only a handful of papers~\cite{nikishov64,ritus85,Voroshilo2010,ilderton11b}.)

A simple estimate suggests that these processes should be negligible unless one of the initial particle species has a density of the order of one particle per Compton wavelength cubed. (This is typical for e.g.~`many to few' processes, and will be made explicit in the formulae below.) This requirement for the species density ($\sim 7\times 10^{28}~\textrm{cm}^{-3}$) is about $10^4$ times denser than solid density ($\sim 10^{24}~\textrm{cm}^{-3}$). However, this should be verified by e.g.~calculation and simulation, and furthermore there are several situations in laser-plasma physics where high particle densities can occur.
For example, at the boundary of an irradiated solid target~\cite{Robinson_2009,Pop2009Schlegel,PRE2011Gonoskov,TangPRE2017,Tang_2019}, an extremely dense electron foil is compressed by ultrarelativistic lasers.
Also, at very high values of the laser field intensity parameter $\xi=eE/m\omega_l \approx O(10^{3})$ (where $e$ and $m$ denote the positron charge and mass and $E$ and $\omega_l$ are the laser electric field amplitude and frequency respectively), QED cascades comprising chains of NLC and NBW processes are predicted to occur \cite{PRL2011Nerush,ridgers12,king13a,bulanov13,jirka16,gonoskov17,PRE2019Efimenko}. In such cascades, electron-positron plasmas are produced and could be compressed to densities much higher than the plasma relativistic critical density.

In this paper we derive a numerical implementation of one-photon pair annihilation and investigate its relevance to the above situations. The paper is organised as follows. In Sec.~\ref{theory}, we calculate the probability of annihilation to one photon, derive its locally constant field approximation (LCFA)~\cite{PRA2019Ilderton} and benchmark against exact results for a circularly-polarised monochromatic field. We then present a study of the dependency of parameters for the probability and a comparison with the background two-photon process. Approximate scaling arguments are also obtained. In Sec.~\ref{simulations} we demonstrate numerical implementations of our results and investigate the relevance of one-photon pair annihilation to laser-plasma and cascade scenarios. We conclude in Sec.~\ref{conclusion}.

\section{One-photon Pair-Annihilation}\label{theory}
\begin{figure}[t!]
\includegraphics[width=0.45\textwidth]{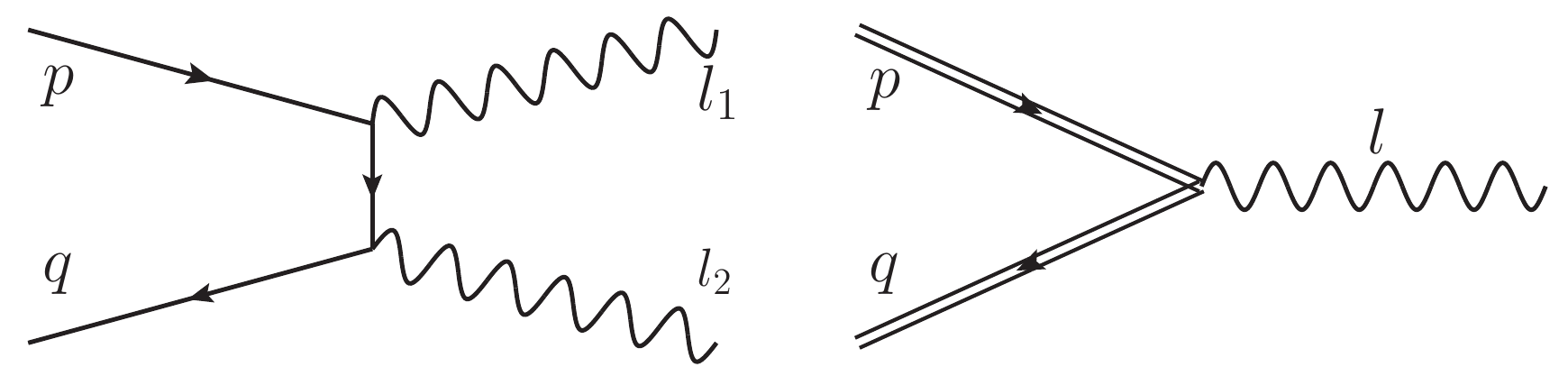}
\caption{Feynman diagram for pair annihilation. Left: Two-photon process in vacuum. right: One-photon process in a field background. Double lines denote dressed Volkov states.}
\label{Fig_Feynman_diagram}
\end{figure}
Electron-positron annihilation in vacuum yields at least two photons~\cite{dirac_1930, greiner2008quantum}. However, in the presence of a background field, annihilation to one photon becomes kinematically accessible; see~Fig.~\ref{Fig_Feynman_diagram}. At low background intensities, the leading-order process is again two-photon emission, but with one of the photons emitted \textit{into} the background. Therefore, when we consider one-photon pair-annihilation, we are also summing over processes that are degenerate with it, such as the (unobservable) emission back to the field~\cite{ilderton2019absorption}.

\begin{figure}[t!]
\includegraphics[width=0.48\textwidth]{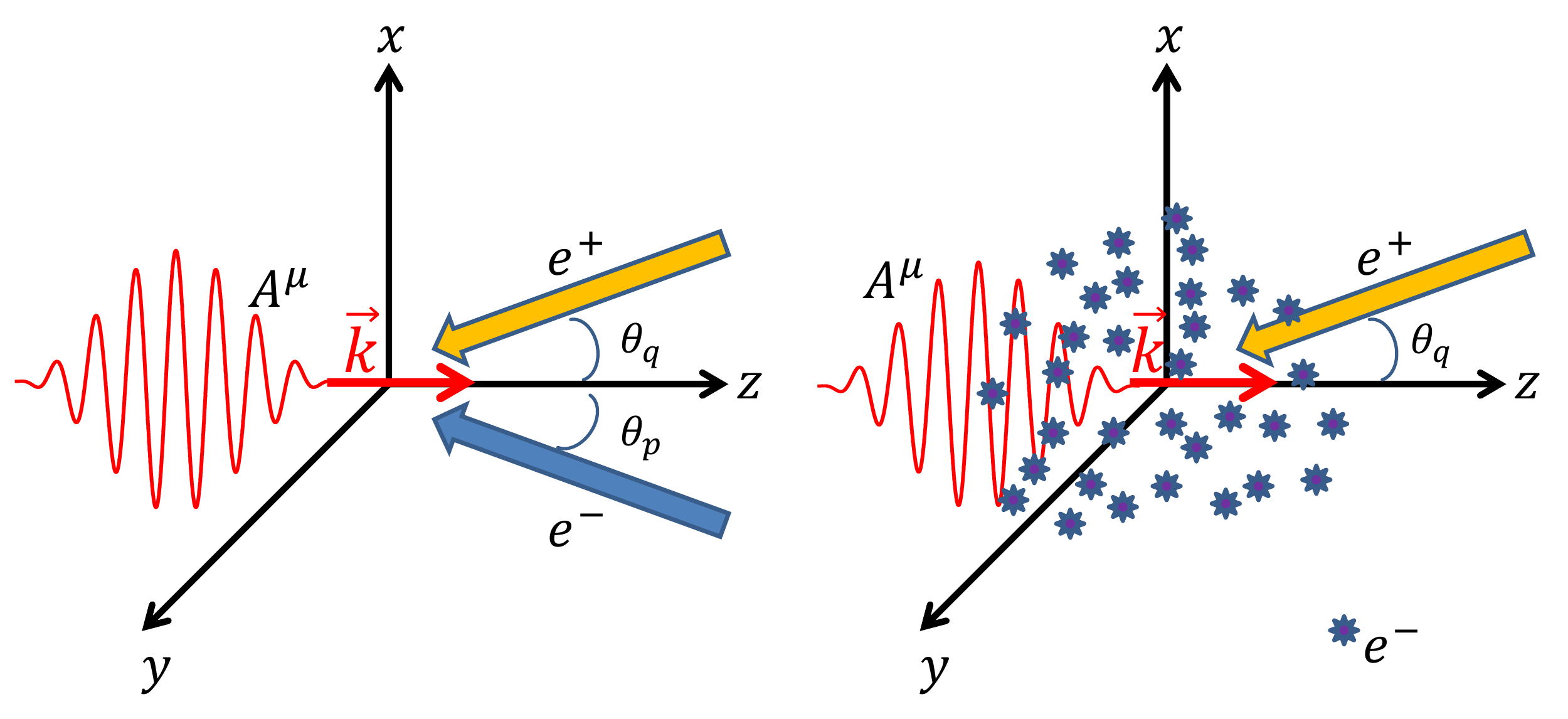}
\caption{Scheme of one-photon pair annihilation. (a) An electron ($e^{-}$) beam and a positron ($e^{+}$) beam collide with a laser pulse ($A^{\mu}$). (b) A positron beam impinges a dense electron target.}
\label{Fig_Setup}
\end{figure}

Here we briefly outline the derivation of one-photon pair-annihilation in pulsed plane wave backgrounds, modelling intense laser pulses. We highlight only those parts of the calculation that differ relative to the more standard $1\rightarrow2$ one-vertex QED processes (a more detailed example of our derivation can be found in Ref.~\cite{PRA2019Ilderton}). We use natural units $\hbar$=$c$=1 throughout and the QED coupling constant is $\alpha=e^2\approx 1/137$. The interaction scenario is set up in Fig.~\ref{Fig_Setup}: we can study the phenomenology of a single process by considering an initial electron (positron) with momentum $p^\mu$ ($q^\mu$) annihilates to a photon with momentum $l^\mu$ in a background laser field as shown in (a). However, an experimental scenario will likely involve high particle density of one species as a target, which we choose to be the electrons, and consider the positrons to be in a beam as shown in (b). The laser field is modelled by the potential $eA^{\mu}(\phi) =a^{\mu}(\phi)= (0,a^1(\phi),a^2(\phi),0)$, in which $\phi=k\cdot x$, with $k=\omega_l(1,0,0,1)$ being the laser wavevector.  The electron is described as a standard Volkov wavefunction~\cite{volkov35},
\begin{align}
\varPsi_{e^{-}}(p)=&\sqrt{\frac{m_0}{Vp^{0}}}\left(1+\frac{\slashed{k}\slashed{a}}{2k\cdot p}\right)u_{p,\sigma}\nonumber\\
                      &\e^{-i\left[p\cdot x+\int^{\phi}d\phi'\left(\frac{p\cdot a}{ k\cdot p}-\frac{a^2}{2k\cdot p}\right)\right]}\,,
\label{Eq_Volkov_state_electron}
\end{align}
where the spinor $u_{p,\sigma}$ satisfies the relation: $\sum_{\sigma}u_{p\sigma}\overline{u}_{p\sigma}=(\slashed{p}+m)/(2m)$. The positron is also described by a Volkov wavefunction, but we include a momentum-space wavepacket $\rho(q)$ to represent a beam of positrons. Writing  $\varPsi_{e^{+}}(q)$ for the positron Volkov wavefunction, the positron is described by
\begin{align}
\Phi_{e^{+}}=\int\frac{d^3q}{(2\pi)^3}\frac{m}{q^0}\rho(q)\varPsi_{e^{+}}(q)\,,
\end{align}
where $\rho$ obeys the normalisation condition
\begin{align}
\int\frac{d^3q}{(2\pi)^3}\frac{m}{q^{0}}\left|\rho(q)\right|^2=1\,.
\label{Eq_wave_packet_normalisation}
\end{align}
The $S$-matrix element for annihilation is
\begin{align}
	\tsf{S}_{\tsf{fi}}&=ie\sqrt{\frac{2\pi}{l^{0} V}} \int d^4x ~\overline{\Phi}_{e^+} \slashed{\epsilon}~\e^{il\cdot x}\, \varPsi_{e^{-}}\,.
\end{align}
where $\epsilon_\mu$ is the polarisation of the produced photon, obeying $\epsilon\cdot \epsilon=-1$ and $l\cdot\epsilon=0$.  The probability $\tsf{P}$ for annihilation is then
\begin{align}
	\tsf{P} = \int \frac{Vd^{3}l}{(2\pi)^3} \, \frac{1}{4}\sum_{\tsf{pol}, \tsf{spin}} \left|\tsf{S}_{\tsf{fi}}\right|^2 \,,
\end{align}
where we sum over the polarisation of the outgoing photon and average over the spins of the incoming pair. For details of the calculation see e.g.~\cite{ilderton2019absorption}. We find:
\begin{align}
\tsf{P}&=\frac{\alpha\lambda_c^3}{8\pi^2 V }\frac{m}{p^{0}} \int\frac{d^3q}{(2\pi)^3}\frac{m}{q^0}~\frac{\left|\rho(\bm{q})\right|^{2}}{\eta_q \eta_{l}}\int d\phi_1 d\phi_2\nonumber\\
     &\left\{1+\frac{\left[a(\phi_1)-a(\phi_2)\right]^2}{m^{2}} \frac{\eta^2_p+\eta^2_q}{4\eta_q \eta_p} \right\}\e^{\frac{-i}{2\eta_{l}}\int_{\phi_2}^{\phi_1}d\phi'\frac{\pi^2_{l}}{m^2}}\,, \label{eqn:pulse1}
\end{align}
where $\eta_p=k\cdot p/m^2$, $\eta_q=k\cdot q/m^2$, $\eta_{l}=\eta_p+ \eta_q$, and $\lambda_c=2\pi/m$ is the electron Compton wavelength. We define the shorthand $\pi_{l}(\phi) = \pi_{p}(\phi)+\pi_{q}(\phi)$, and $\pi^{\mu}_{p}=p^{\mu}+a^{\mu}-(2p\cdot a+a^2)k^{\mu}/(2k\cdot p)$ ($\pi^{\mu}_q=q^{\mu}-a^{\mu}+(2q\cdot a-a^2)k^{\mu}/(2k\cdot q)$) is the instantaneous four-momentum of the electron (positron) in a plane-wave. Note that the probability contains the leading density factor $\lambda^3_c/V$.

\subsection{LCFA}

To derive the LCFA, we follow the usual procedure of rewriting the external-field phases in terms of an average phase $\psi = (\phi_1 + \phi_2)/2$ and an interference phase $\vartheta = \phi_1-\phi_2$~\cite{PRA2019Ilderton}, expanding the exponent to order $\vartheta^{3}$:
\bea
 \int_{\phi_1}^{\phi_2}d\phi^{\prime}\pi_{l}^{2} \to \vartheta \pi_{l}^{2}(\psi) + \frac{\vartheta^{3}}{24}[\pi_{l}^{2}(\psi)]'',
\eea
and the pre-exponent up to order $\vartheta^{2}$ through the replacement  $[a(\phi_1)-a(\phi_2)]^{2} \to -m^{2}\vartheta^{2}\bm{\xi}^{2}(\psi)$, where we define the normalised electric field $\bm{\xi}$ through $a'/m = (0,\bm{\xi})$. This allows us to integrate \eqnref{eqn:pulse1} over $\vartheta$. The probability becomes
\begin{align}
	\tsf{P}&=\frac{\alpha\lambda^3_c}{2\pi V}\int\frac{d^3q ~m}{(2\pi)^3 q^0}\left|\rho(q)\right|^{2} g(p,q,\bm{\xi})\,,
\label{Eq_LCFA_wave_packet}
\end{align}
which has the form of an incoherent average over the positron wavepacket $|\rho(q)|^2$ and the probability for one-photon annihilation of a pair with \textit{definite} momenta $p$ and $q$, which is encoded in $g$. The dependence on the particle momenta and the field $\bm{\xi}$ is described by
\begin{align}
g(p,q,\bm{\xi})=\frac{m}{p^{0}\eta_q}\int d\psi~f(p,q,\bm{\xi})\,,
\label{Eq_LCFA_wave_packet_integral_phase}
\end{align}
in which
\begin{align}
f(p,q,\bm{\xi})&= \left(\frac{\chi^{1/3}_{q}\chi^{1/3}_{p}}{\chi^{4/3}_{l}} +\frac{\chi^2_p+\chi^2_q}{\chi^2_{l}}z\right)\textrm{Ai}(z)\,,
\label{Eq_LCFA_Integrand_F}
\end{align}
where all $\chi$ variables depend on the average phase $\psi$ via $\chi_p=\eta_p|\bm{\xi}(\psi)|$, $\chi_q=\eta_q|\bm{\xi}(\psi)|$, $\chi_{l}=\chi_{p}+\chi_{q}$ and $\textrm{Ai}(z)$ is the Airy function with argument
\bea
z=\frac{(\pi_{p}+\pi_{q})^2}{m^{2}\,\chi_{l}}\left(\frac{\chi_{q}\chi_{p}}{\chi_{l}}\right)^{\frac{1}{3}}\,.
\eea
Observe that the LCFA result depends not only on the quantum nonlinearity parameter $\chi_{p,q}$, but also on the local momenta of the two particles, $\pi_{p,q}$. 

\subsection{LCFA Benchmarking}
To benchmark the LCFA result, we consider one-photon pair-annihilation in a circularly polarised monochromatic field: $\bm{\xi}(\psi)=\xi[\cos\psi,-\sin\psi,0]$. Unlike the case of $1\to2$ processes, where the LCFA can be compared straightforwardly, in $2\to1$ processes the number of outgoing momentum integrals is not always sufficient to evaluate all momentum-conserving delta-functions. However, if one of the incoming particles is in a wavepacket state, as we consider here, then all the delta-functions can be evaluated and the LCFA can again be benchmarked straightforwardly.

We take our positron wavepacket to be
\begin{align}
\left|\rho(q)\right|^2=(2\pi)^3\frac{q^{-}}{m}\nu(q^{-}) \frac{4\,\textrm{ln}(2)}{\pi\Delta^2 m^2}\e^{-\frac{4\,\textrm{ln}(2)}{\Delta^2 m^2}\left|\bm{q}^{\perp}-\bm{q}^{\perp}_{i}\right|^2}\,,
\end{align}
which is Gaussian distributed in transverse momentum with full width at half maximum $\Delta m$, while the longitudinal wavepacket $\nu(q^{-})$ satisfies the normalisation condition: $\int^{\infty}_{0} dq^{-}\nu(q^{-}) =1$. This ansatz for the wavepacket facilitates the intended comparison by matching well with the symmetries of the plane wave background. The explicit form of $\nu(q^-)$ will not be required, as we will focus on transverse momentum dependence.

Inserting the above wavepacket into Eq.~(\ref{Eq_LCFA_wave_packet}), we can obtain the probability:
\begin{align}
\tsf{P}^{\tsf{lcfa}}_{\nu}&=\frac{32\pi^2\textrm{ln}(2)\delta(0)\alpha}{V k^{0}p^{0}\Delta^2} \int^{1}_{0}\frac{dv~\nu(q^{-})}{(1-v)^2}~h_{l}(v)\,,
\label{Eq_LCFA_wave_packet_circular}
\end{align}
where $\delta(0)=\int d\psi/(2\pi)$,  $v=\eta_{q}/\eta_{l}$ and
\begin{align}
h_{l}(v)&=\frac{1}{2\pi}\int d^2\bm{r}\int^{2\pi}_{0} d\psi~\e^{-\frac{4\textrm{ln}(2)}{\Delta^2u}\left|\bm{r}-\bm{r}_{i}\right|^2}\nonumber\\
&~~~\times v^2\left[\frac{u}{(v\chi_{p})^{2/3}} +(1+u^2)z\right]\textrm{Ai}(z)\,,
\label{Eq_LCFA_wave_packet_circular1}
\end{align}
with the definitions:
\begin{align}
\bm{r}&=\frac{\bm{q}^{\perp}}{m}\sqrt{u}-\frac{\bm{p}^{\perp}}{m}\sqrt{\frac{1}{u}}\,,~ \bm{r}_{i}=\frac{\bm{q}^{\perp}_{i}}{m}\sqrt{u}-\frac{\bm{p}^{\perp}}{m}\sqrt{\frac{1}{u}}\,,\nonumber
\end{align}
where $u=\eta_p/\eta_q$.

\begin{figure}[t!]
\includegraphics[width=0.48\textwidth]{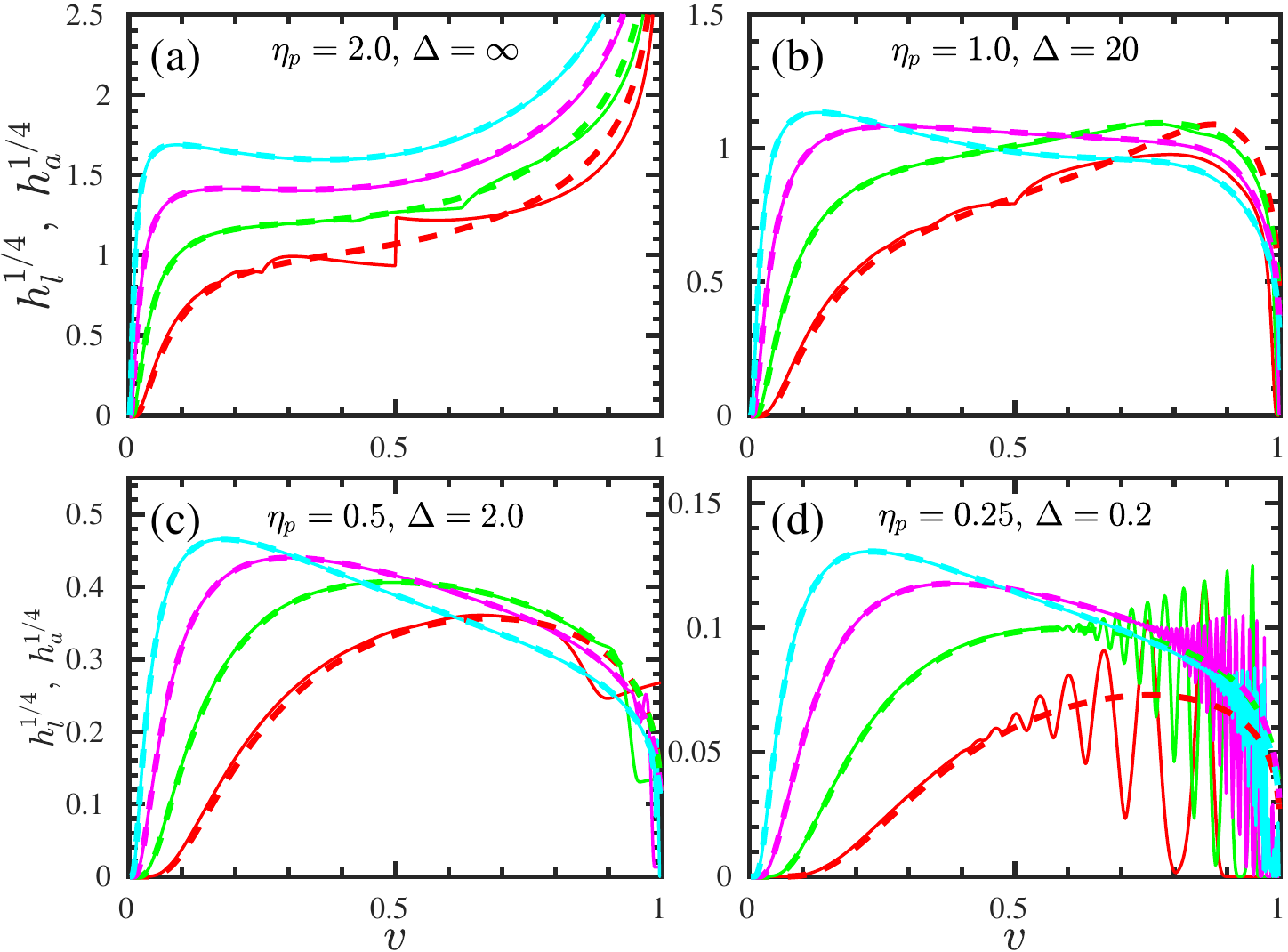}
\caption{LCFA \textit{vs} exact result. (a) $\Delta=\infty$, $\eta_p=2$. (b) $\Delta=20$, $\eta_p=1$. (c) $\Delta=2$, $\eta_p=0.5$. (d) $\Delta=0.25$, $\eta_p=0.2$. Dashed (Solid)lines are for LCFA (exact) result. Red lines: $\xi=1$; Green lines: $\xi=2$, Magenta lines: $\xi=4$, and Blue lines: $\xi=8$. In (b), (c), (d), $q_{i,x}=-p_{x}=\xi$ and $q_{i,y}=p_{y}=0$.}
\label{Fig_LCFA_benchmark}
\end{figure}

We find the annihilation probability in a circularly polarised monochromatic wave to be:
\begin{align}
	\tsf{P}^{\tsf{mono}}_{\nu}&=\frac{32\pi^2\textrm{ln}(2)\delta(0)\alpha}{Vk^0p^{0}\Delta^2} \int^{1}_{0} \frac{d v~\nu(p^{-})}{(1-v)^2}~h_{a}(v)\,,
\label{Eq_cir_full_packet0}
\end{align}
in which $h_a$ contains a sum over harmonics
\begin{align}
h_{a}(v)=\sum_{n\geq n_{v}}^{\infty}\textrm{T}_{n}
\label{Eq_cir_full_packet1}
\end{align}
with the lower harmonic bound $n_{v}=(1+\xi^2)/(2\eta_{p}v)$ and
\begin{align}
	\textrm{T}_{n}(v)=\int^{\pi}_{-\pi} d\varphi~\e^{-\frac{4\textrm{ln}(2)}{\Delta^2 u}(\bm{r}_{n}-\bm{r}_{i})^2}\textrm{H}_{n}
\label{Eq_cir_harmonic}
\end{align}
where $\vphi$ is the angle between $\bm{r}_{n}$ and $\bm{r}_{i}$, $\cos(\vphi)={\bm{r}_{n}\cdot\bm{r}_{i}}/(r_{n}r_{i})$, $r_i:=|\bm{r}_i|$ is a function of only initial variables and
\[
 r^2_{n}=|\bm{r}_{n}|^{2}=2n\eta_{p}\frac{v-v^{*}}{v(1-v)}\,,\quad v^{*}=\frac{1+\xi^2}{2n\eta_p},
\]
and $n$ is the harmonic number
\begin{align}
\textrm{H}_{n}&=\frac{\xi^2}{4}\left[\frac{n^2-s_{n}^2}{s_{n}^2}J^2_{n}(s_{n})+J^{\prime\,2}_{n}(s_{n})\right]\frac{1-2v^2u}{v^2u} +\frac{J^2_n(s_{n})}{2}\,,\nonumber
\end{align}
where $J_{n}(s_{n})$ is the Bessel function of the  first kind, with argument
\[
s_{n}=\frac{\xi^2\sqrt{2n\eta_{p}(v-v^{\ast})}}{\chi_{p}v}
\]
We compare in Fig.~\ref{Fig_LCFA_benchmark} the LCFA result Eq.~(\ref{Eq_LCFA_wave_packet_circular1}) with the exact monochromatic result Eq.~(\ref{Eq_cir_full_packet1}), for various parameters and wavepackets of different widths. In the figure, the LCFA is represented by dashed lines and the monochromatic result by solid lines.

In Fig.~\ref{Fig_LCFA_benchmark} (a), the LCFA result for a flat wavepacket matches well with the exact calculation when the field intensity is relatively strong; notice that the LCFA cannot reproduce the low-$n$ harmonic structure visible at e.g.~$\xi=1$ (red lines), as expected from other investigations of the LCFA~\cite{PRA2015Harvey}, see also~\cite{meuren17,PRA2019Ilderton,king2019uniform}. For pair-annihilation, we find a similar dependency on the minimum harmonic as for the time-reversed process of NBW pair-creation~\cite{nikishov64}. There is a \textit{lower bound} $n_v$ on the harmonic number which \textit{increases} with intensity; as the LCFA does better at reproducing results where large numbers of higher-harmonics contribute~\cite{PRA2015Harvey,meuren17,PRA2019Ilderton,king2019uniform}, the quality of the LCFA improves quickly here, being extremely accurate already for $\xi=4$ (magenta lines). Consequently, a similar effect results from decreasing $\eta_p$: this also raises the harmonic lower bound, leading to weaker harmonic structure, meaning that the LCFA gives a better approximation of the monochromatic result even at low laser intensities. This is confirmed in Fig.~\ref{Fig_LCFA_benchmark} (b) and (c).

\begin{figure}[t!]
\includegraphics[width=0.45\textwidth]{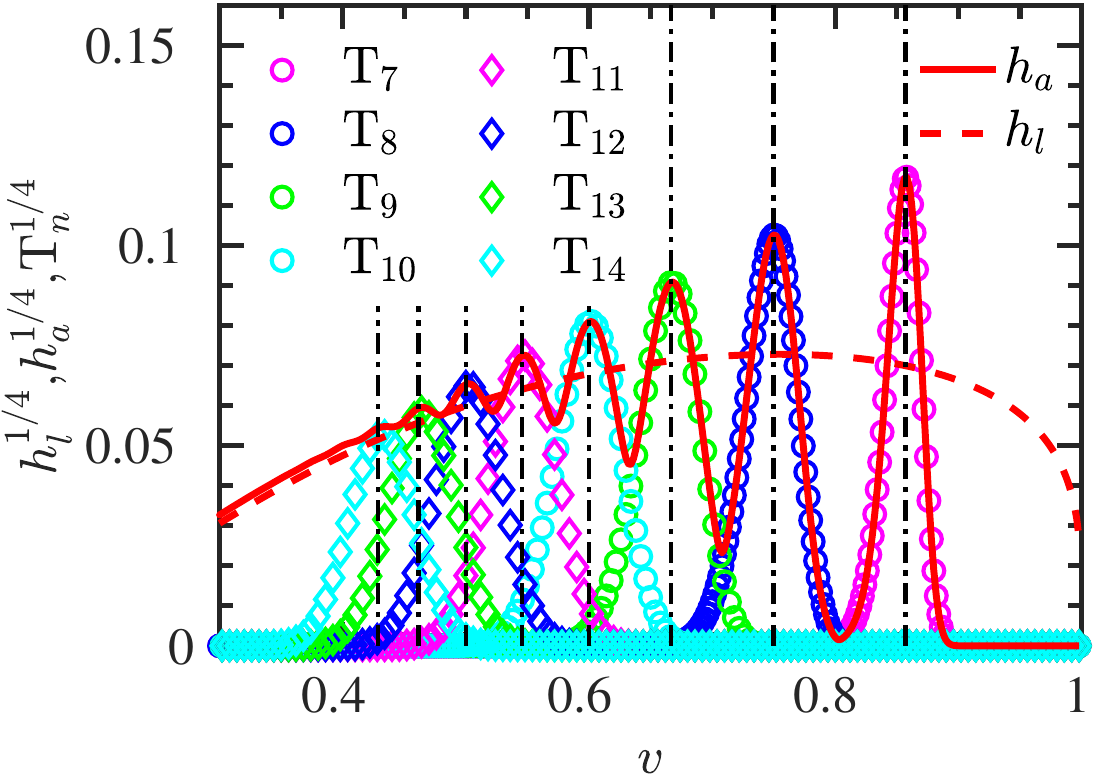}
\caption{Harmonic peak structure. We zoom in the peak structure of the red line for $\xi=1$ in Fig.~\ref{Fig_LCFA_benchmark} (d), and separate the contribution from different order of harmonics (cycles and diamonds). The black dash-dotted lines denote $v_n$, as in Eq.~(\ref{Eq_location_of_harmonic}), corresponding to each harmonic.}
\label{Fig_Harmonic}
\end{figure}

We also highlight the behaviour at $v=1$ in Fig.~\ref{Fig_LCFA_benchmark} (a) for the flat wavepacket, where $h_l,h_a\to\infty$. This divergence comes from the superposition of an infinite number of states with the same longitudinal momentum. However, this behaviour is different when the included wavepacket has a finite momentum bandwidth. As we compare, in order, Figs.~\ref{Fig_LCFA_benchmark}~(a)-(d), the wavepacket becomes narrower, and the exact result oscillates rapidly as $v \rightarrow 1$, with the oscillating structure spreading to lower $v$ as the wavepacket narrows. These rapid oscillations result from an interplay between the harmonics and the Gaussian wavepacket: the contribution of each harmonic is effectively localised by the narrow wavepacket. The dominant contribution from each harmonic originates from the condition $(\bm{r}_i-\bm{r}_{n})^2=0$, as can be seen from the exponent in Eq.~(\ref{Eq_cir_harmonic}). This can be solved for $v$, showing that the $n^\text{th}$ harmonic $\tsf{H}_n$ will be restricted to contribute around $v\simeq v_n$ where
\begin{align}
	v_{n} :=\frac{1+2\xi^{2}}{2\eta_p n} \;.
\label{Eq_location_of_harmonic}
\end{align}
We have used here that $\bm{p}^{\perp}=-\bm{q}^{\perp}_i$ and  $|\bm{p}^{\perp}|=\xi$ as in Fig.~\ref{Fig_LCFA_benchmark}. To see these effects explicitly we zoom in to the peak structure for $\xi=1$ (red solid line) in Fig.~\ref{Fig_LCFA_benchmark} (d) and highlight, in  Fig.~\ref{Fig_Harmonic}, the contribution from different harmonics. As predicted, the $n^\text{th}$ harmonic is highly localised around the point $v = v_n$. Also, the harmonics that significantly contribute are substantially above the lower bound $n_v$. It is clear from Fig.~\ref{Fig_Harmonic} that the separation of the harmonic contributions, due to the wavepacket, is responsible for the appearance of the oscillatory structure as the wavepacket width decreases. Furthermore, the LCFA result fails to manifest this peak structure at all. As $v$ decreases, the harmonic peaks overlap and can be matched better by the LCFA, but this agreement is again lost as the wavepacket continues to narrow and the harmonic peaks become much sharper, as in Fig.~\ref{Fig_LCFA_benchmark} (c) and (d). (We also find the oscillation frequency increases with the increase of the laser intensity,  Fig.~\ref{Fig_LCFA_benchmark} (d).) We conclude that the LCFA is unable to reproduce the \BK{$2\to1$} physics of narrow (momentum space) wavepackets.

\subsection{Phenomenology}
In this section, we study the dependence of one-photon pair-annihilation on the incident particle parameters, assuming plane-wave initial states. The particle momenta are expressed in spherical polar co-ordinates as depicted in Fig.~\ref{Fig_Setup} (a): $\bm{p}=-(E^2_p-m^2)^{1/2}[\sin\theta_p\,\cos\vphi_p,\sin\theta_p\,\sin\vphi_p,\cos\theta_p]$, and $\bm{q}=-(E^2_q-m^2)^{1/2}[\sin\theta_q\,\cos\vphi_q,\sin\theta_q\,\sin\vphi_q,\cos\theta_q]$ where $E_{p},~\theta_p,~\vphi_p$ ($E_{q},~\theta_q,~\vphi_q$) are the incident energy, polar and azimuthal angle of electron (positron). We analyse one cycle of a monochromatic field with i) linear polarisation: $a^{\mu}(\psi)=m\xi[0,\cos\psi,0,0]$, and ii) circular polarisation: $a^{\mu}(\psi)=m\xi[0,\sin\psi,\cos\psi,0]$. For linear polarisation, we consider a head-on collision with the laser background ($\vphi_p=\vphi_q=0$).

It is helpful for what follows to understand where the dominant contributions to $f(p,q,\bm{\xi})$ in Eq.~(\ref{Eq_LCFA_Integrand_F}) come from, in terms of phase $\psi$ and as a function of the particle momenta. The pair should have similar energy $E_p\approx E_q$, and we find that $f(p,q,\bm{\xi})$ exhibits one (two) sharp peaks per laser cycle for circular (linear) polarisation. These peaks appear at the points where $\pmb{\pi}_p(\psi)$ is parallel to $\pmb{\pi}_{q}(\psi)$. For linear polarisation: if $\theta_p=\theta_q$, the peaks appear at the points where $a(\psi)=0$, and if $\theta_p=-\theta_q=\theta$, the peaks appear at $m\xi\cos\psi=E_p\sin\theta$.
For circular polarisation: a peak appears at $\bm{a}(\psi)=E_p[\sin\theta\,\cos\psi,\sin\theta\,\sin\psi,0]$, if $\bm{p}^{\perp}+\bm{q}^{\perp}=0$. Because the EM field rotates in a circularly-polarised background, these acceptance peaks are much narrower than for a linearly-polarised background. Thus one-photon pair-annihilation is much more effective in a linearly polarised laser.

\begin{figure}[t!!]
\includegraphics[width=0.48\textwidth]{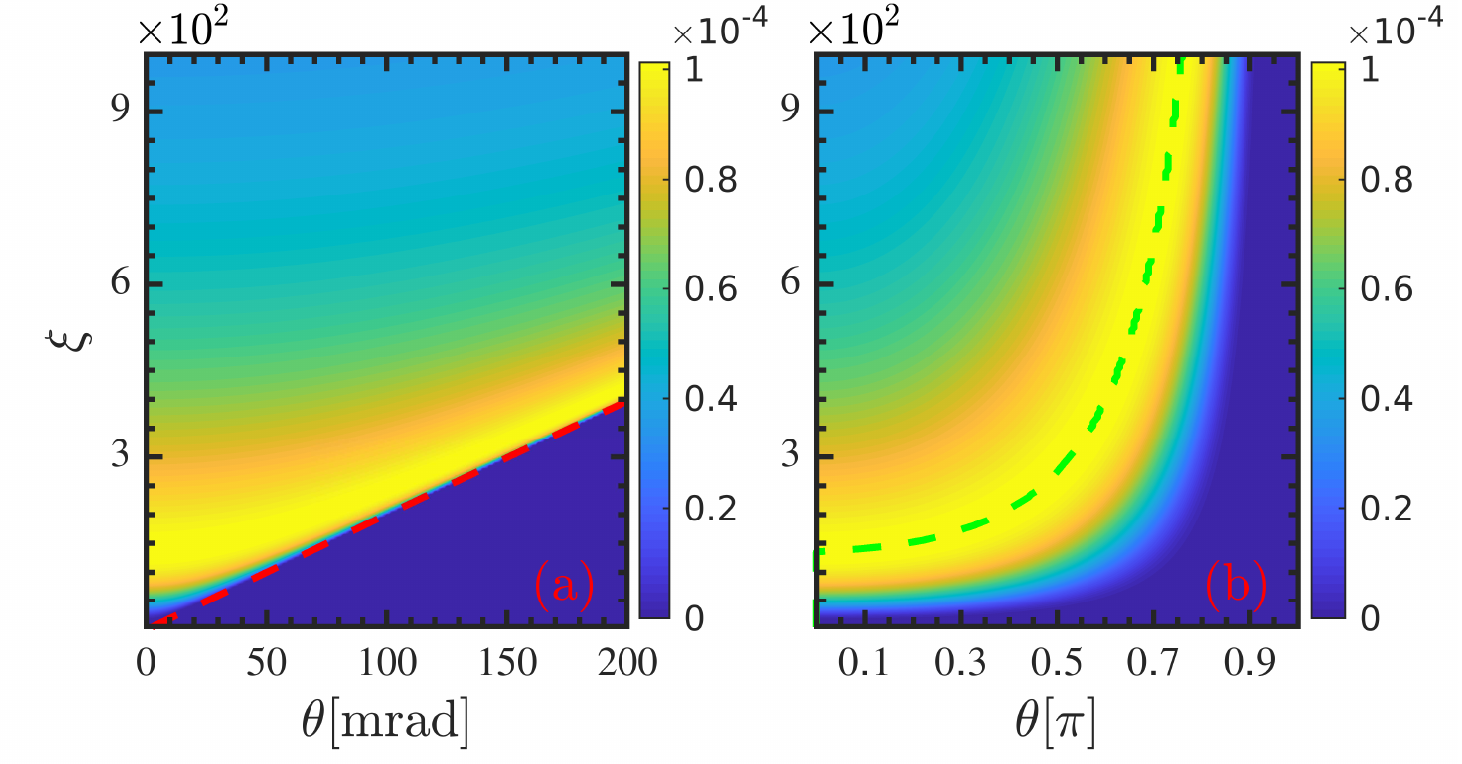}
\caption{Parametric dependency of $g(p,q,\xi)$ on the laser amplitude $\xi$ and the particles' incident angle $\theta_p$, $\theta_q$.
(a) $\theta_p=-\theta_q=:\theta$. (b) $\theta_p=\theta_q=:\theta$.
The red dashed line in (a) is $\xi=|\bm{p}|\sin(\theta)/m$, and the green dashed line in (b) corresponds to $\eta_p\xi=4/3$.
The particle energy is fixed $E_p=E_q=2000m$, and a laser is linearly polarised, with frequency $\omega_l=1.24~\textrm{eV}$.}
\label{Fig_wave_packet_inte_F1}
\end{figure}

Fig.~\ref{Fig_wave_packet_inte_F1} shows the dependency of the integrated expression $g(p,q,\bm{\xi})$  from \eqnref{Eq_LCFA_wave_packet_integral_phase}, on the laser amplitude $\xi$ and the particles' incident angle $\theta$.
In Fig.~\ref{Fig_wave_packet_inte_F1} (a), we can see that, in order that the process is not strongly suppressed, the laser intensity must be increased for larger values of incident collision angle $\theta$. The reason for this, is that the laser field must be strong enough to make the local momenta of the pair particles, $\pmb{\pi}_{p}$ and $\pmb{\pi}_{q}$, parallel to one another. If the pair particles propagate parallel to one another and collide head-on with the laser pulse, there is an optimal value of $\xi$, above which the probability then decreases. This is because, even though the strength of the interaction is increased, there is a suppression at high intensities due to a narrowing of the effective phase width in the integrand $f(p,q,\bm{\xi})$. For a high enough intensity, the most probable set-up for one-photon pair-annihilation is actually when the collision is not directly head-on, as shown in Fig.~\ref{Fig_wave_packet_inte_F1} (b). This is because one-photon pair-annihilation achieves the largest probability if the quantum nonlinearity parameter $\chi=4/3 \sim [1+\cos(\theta)]\xi$ (this will be further commented on in the approximations section and Eq.~(\ref{Eq_approximation3})).

\begin{figure}[t!]
\includegraphics[width=0.49\textwidth]{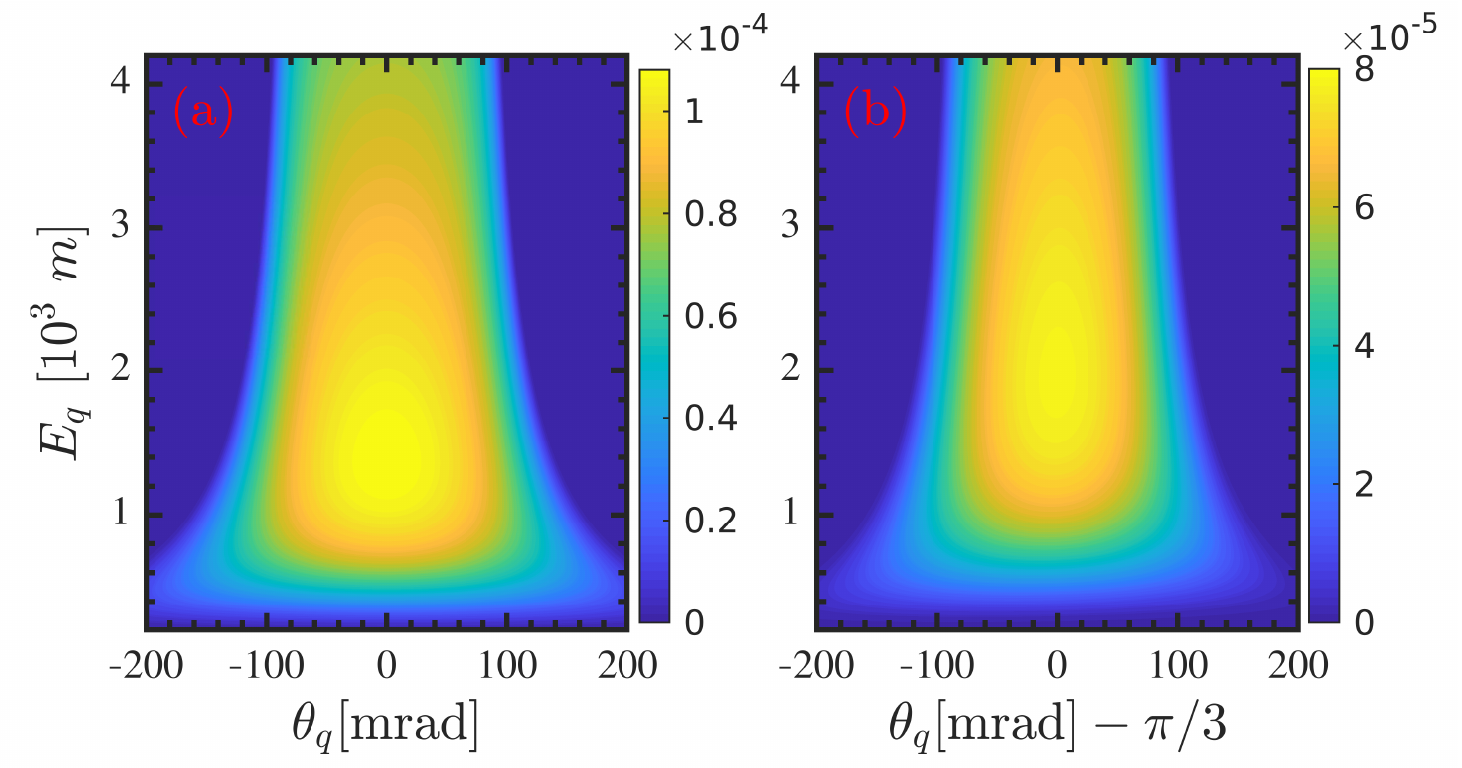}\\
\caption{Parametric dependency of $g(p,q,\bm{\xi})$ on the incident parameters of the positron. (a) $\theta_p=0$, $E_p=1360m$. (b) $\theta_p=\pi/3$, $E_p=1360m$. The laser amplitude is $\xi=100$, and the other parameters are same in Fig.~\ref{Fig_wave_packet_inte_F1}.}
\label{Fig_wave_packet_inte_F2}
\end{figure}

In Fig.~\ref{Fig_wave_packet_inte_F2}, we show the dependency of $g(p,q,\bm{\xi})$ on the incident parameters ($E_q$, $\theta_q$) of the positron for the given laser amplitude $\xi=100$ and electron incident parameters [$E_p=1360m$, $\theta_p=0$ in (a) and $\theta_p=\pi/3$ (b)]. As we can see, the largest probability could be obtained if the pair particles have the same initial parameters ($\theta_p\approx \theta_q$, $E_p \approx E_q$). With a relative larger incident angle in Fig.~\ref{Fig_wave_packet_inte_F2} (b), the process would be less effective because $\chi_p\sim 1+\cos(\theta)$ is smaller.

\begin{figure}[t!]
\includegraphics[width=0.49\textwidth]{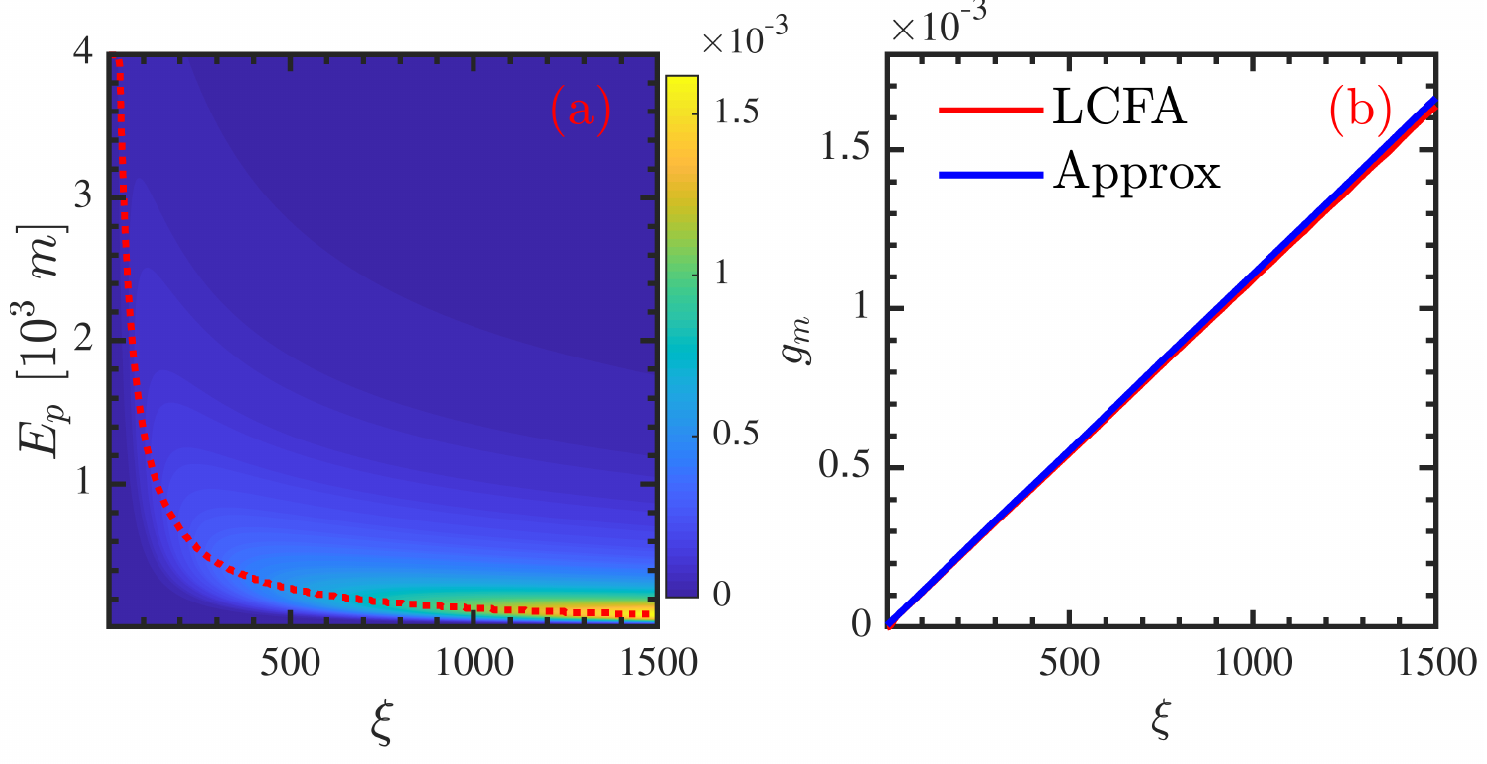}
\caption{(a) Parametric dependency of $g(p,q,\xi)$ on the particle energy $E_p=E_q$ and laser amplitude $\xi$. The red dotted line denotes the particle energy giving the largest $g(p,q,\xi)$. (b) Largest $g(p,q,\xi)$ for a given laser intensity $\xi$. The red line corresponds to the red dashed line in (a) and the blue lines comes from the approximation Eq.~(\ref{Eq_approximation_optimal}).
Head on collision ($\theta_p=\theta_q=0$) is applied, and the other parameters are same in Fig.~\ref{Fig_wave_packet_inte_F1}.}
\label{Fig_wave_packet_inte_F3}
\end{figure}

In Fig.~\ref{Fig_wave_packet_inte_F3}, we consider the dependency of $g(p,q,\bm{\xi})$ on the particle energy and laser amplitude with head-on collisions $\theta_p=\theta_q=0$.
As shown in Fig.~\ref{Fig_wave_packet_inte_F3} (a), a stronger laser field could induce larger annihilation probability and also decrease the requirement for the particle energy (see the red dotted line). This is because the probability has a maximum at  $\chi_p=2/3$ [this will be discussed further in Eq.~(\ref{Eq_approximation_optimal})], and in Fig.~\ref{Fig_wave_packet_inte_F3} (b), the maximal probability for a given laser field $\xi$ increases with a stronger laser field.

\subsection{Approximations}
To understand the dependency of one-photon pair-annihilation on experimental parameters, it is useful to approximate the phase integral in Eq.~(\ref{Eq_LCFA_wave_packet_integral_phase}).

To simplify the calculation, in this section we consider two cases, corresponding to
Fig.~\ref{Fig_wave_packet_inte_F1} (a) and (b) respectively. First of all we assume a head-on collision $\theta_p=\theta_q=0$ (Fig.~\ref{Fig_wave_packet_inte_F1} (a)) with a linearly-polarised monochromatic laser field $\bm{\xi}(\psi)=\xi[\sin(\psi),0,0]$, and consider an integration over one cycle of this field. The argument, $z$, of the Airy function, becomes:
\begin{align}
z=z_{m}[1+\xi^2\cos^{2}(\psi)]\sin^{-2/3}(\psi)\,,
\end{align}
where $z_{m}=\left(\eta_p+\eta_q\right)^{2/3}/(\eta_q\eta_p \xi)^{2/3}$. The leading contribution comes when $a(\psi) \approx 0$ and $z$ is at a minimum (and the Airy functions are at their maximum). If we Taylor-expand $z$ in $\psi$ to order $\psi^{2}$ around the corresponding points at $\psi=\pi/2, ~3\pi/2$, we can integrate over $\psi$ in Eq.~(\ref{Eq_LCFA_wave_packet_integral_phase}) and arrive at:
\begin{align}
g(p,q,\xi)&\approx \frac{2m}{p^{0}\eta_q} \left(\eta_{p}+\eta_{p}\right)^{-\frac{4}{3}}(\eta_p\eta_q)^{\frac{1}{3}} \xi^{-\frac{5}{3}}\frac{\pi}{\sqrt{z_{m}}}\nonumber\\
     &~\left[\left(1 +\Gamma-\frac{1+2\Gamma}{6\xi^2_{0}}\right)2^{\frac{2}{3}}\textrm{Ai}^2\left(2^{-\frac{2}{3}}z_{m}\right)\right.\nonumber\\ &~\left.+\frac{1+2\Gamma+6\xi^2_{0}\Gamma}{3\xi^2_{0}z_{m}}2^{\frac{1}{3}}\textrm{Ai}'^2\left(2^{-\frac{2}{3}}z_{m}\right) \right]\,,
\label{Eq_approximation1}
\end{align}
where $\Gamma = \Gamma(p,q)=(\eta_p^2+\eta_q^2)/(2\eta_q\eta_p)$. If $2^{-\frac{2}{3}}z_{m}\gg1$, we can obtain
\begin{align}
g(p,q,\xi) \approx \frac{m}{p^{0}\eta_q}\frac{\eta_{p}\eta_{q}}{(\eta_{p}+\eta_{q})^2}\frac{1+2\Gamma }{\xi}\exp\left(-\frac{2}{3}z_{m}^{\frac{3}{2}}\right)\,,
\label{Eq_approximation2}
\end{align}
and if $p = q$, $g(p,q,\xi)$ can be further simplified:
\begin{align}
g(p,q,\xi)\approx \frac{3m}{4p^{0}}\frac{1}{\chi_{m}}\exp\left(-\frac{4}{3\chi_{m}}\right)\,,
\label{Eq_approximation3}
\end{align}
where $\chi_{m}=\eta_p\xi$. (This is reminiscent of the famous $\exp(-8/3\chi_{l})$ scaling of the time-reversed process of NBW pair-creation in a constant crossed field in the asymptotic limit $\chi_{l} \ll 1$.)

We can perform the same analysis for $\theta_{p}=\theta_{q}=\theta$ (Fig.~\ref{Fig_wave_packet_inte_F1} (b)). Using the same approximation as in Eq.~(\ref{Eq_approximation3}), we see that, for a given particle energy $E_p=E_q$, $g(p,q,\xi)$ has a maximum if $\chi_{m}=4/3$ (corresponding to the green dashed line in Fig.~\ref{Fig_wave_packet_inte_F1} (b)), and for a given laser amplitude $\xi$, $g(p,q,\xi)$ has a maximum if $\chi_{m}=2/3$, (see the blue line in Fig.~\ref{Fig_wave_packet_inte_F3} (b)). $g(p,q,\xi)$ then takes the value:
\begin{align}
g_{\tsf{m}}=\frac{27}{8}\frac{\omega_l\xi}{m}e^{-2}\,,
\label{Eq_approximation_optimal}
\end{align}
(where we have made use of the relation $k\cdot p\approx 2\omega_lp^{0}$ if $p^{0}\gg1$ in a head-on collision).

The approximations above are for a single cycle of a monochromatic background, but the same approximation can be made for longer pulses by summing over contributions where $a(\psi_i)=0$:
\begin{align}
g(p,q,\xi)&\approx\frac{m}{2p^{0}}\frac{\eta_{p}}{(\eta_p+\eta_q)^2} \sum_{i}\frac{1+2\Gamma}{|\bm{\xi}(\psi_i)|}\e^{-\frac{2}{3}z^{3/2}_{m}(\psi_i)}\,.
\label{Eq_rate_approx_pulse2}
\end{align}

\begin{figure}[t!]
\includegraphics[width=0.48\textwidth]{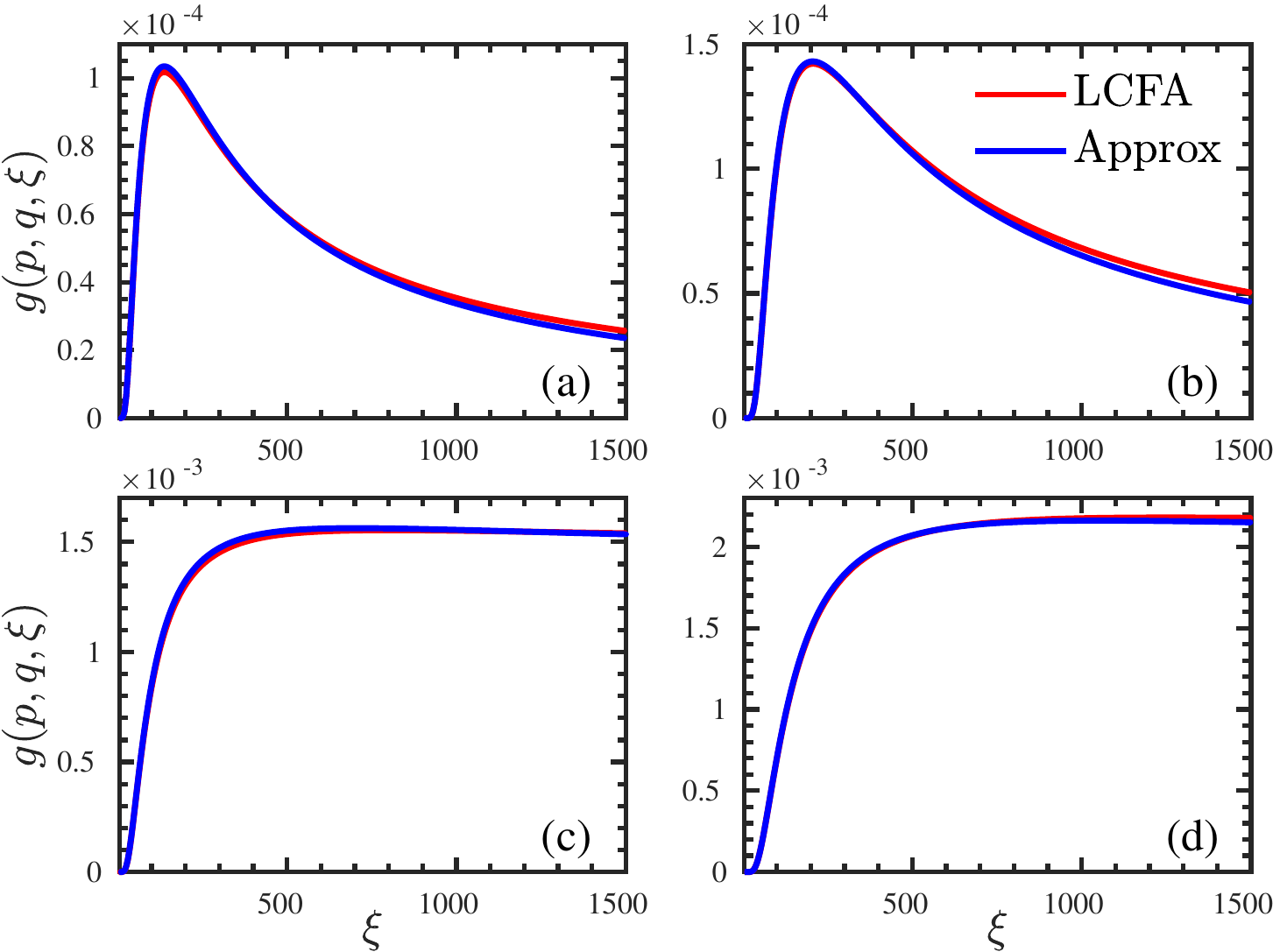}
\caption{Comparison between the numerical calculation of Eq.~(\ref{Eq_LCFA_wave_packet_integral_phase}) and the approximation Eq.~(\ref{Eq_approximation2}) for a head-on collision. In (a) and (b), the results for one cycle of a monochromatic laser pulse $\bm{\xi}(\psi)=\xi~[\sin(\psi),0,0]$ are displayed. In (c) and (d) the cases of a long laser pulse $\bm{\xi}(\psi)=\xi~[\sin(\psi),0,0]~\textrm{sech}^2[\psi/(\omega_{l}T)]$, $\omega_l=1.24~\textrm{eV}$, $T=5T_l$, $T_l=2\pi/\omega_l$ are presented. In (a) and (c), $E_{p}=E_{q}=2000m$; in (b) and (d), $E_p=2000m$, $E_{q}=1000m$.}
\label{Fig_Approximation}
\end{figure}

To demonstrate the validity of the approximation, we show in Fig.~\ref{Fig_Approximation} the comparison between the numerical calculation of Eq.~(\ref{Eq_LCFA_wave_packet_integral_phase}) and the approximation Eq.~(\ref{Eq_approximation2}). As we can see, the approximation works well in a broad parameter region, with the discrepancy growing in the extremely high field and high energy region as $z_m\gtrsim 1$.

\subsection{Comparison with zero-field two-photon pair-annihilation}
Based on Eq.~(\ref{Eq_LCFA_wave_packet}) and the definition of the cross section, $\sigma=(1/|v_{rel}|n_{e^{-}})d\tsf{P}/dt$~\cite{landau2013classical}, where $t$ is time and $|v_{rel}|=(p^{0}q^{0})^{-1}\sqrt{(p\cdot q)^2-m^4}$ is the relative velocity between the pair particles, we can easily calculate the cross section for one-photon pair-annihilation:
\begin{align}
\sigma_1=\frac{2\alpha \lambda_c^2}{\sqrt{\kappa^2-4\kappa}}\frac{1}{2\pi N_{l}}\int d\psi~f(p,q,\bm{\xi})\,,
\label{Eq_annihilation_cross_section_one}
\end{align}
where $\kappa=(p+q)^2/m^2$ is the scaled Mandelstam invariant, $N_{l}$ is the number of laser cycles and we replace the volume factor $1/V$ in Eq.~(\ref{Eq_LCFA_wave_packet}) with the electron density $n_{e^{-}}$. (Here, an ``evening-out'' of the instantaneous cross section is performed by averaging over the phase of the incident laser pulse.) The cross section for the two-photon annihilation process in vacuum is calculated in Ref.~\cite{greiner2008quantum}:
\begin{align}
\sigma_2=&\frac{\alpha^2 \lambda^2_c}{2\pi}\frac{1}{\kappa-4} \left[-\frac{\kappa+4}{\kappa}\sqrt{\frac{\kappa-4}{\kappa}}\right.\nonumber\\
       &\left.+\textrm{ln}\left(\frac{\kappa-2}{2}+\sqrt{\frac{\kappa^2}{4}-\kappa}\right)\frac{\kappa^2+4\kappa-8}{\kappa^2} \right]\,.
\end{align}

\begin{figure}[h]
\includegraphics[width=0.48\textwidth]{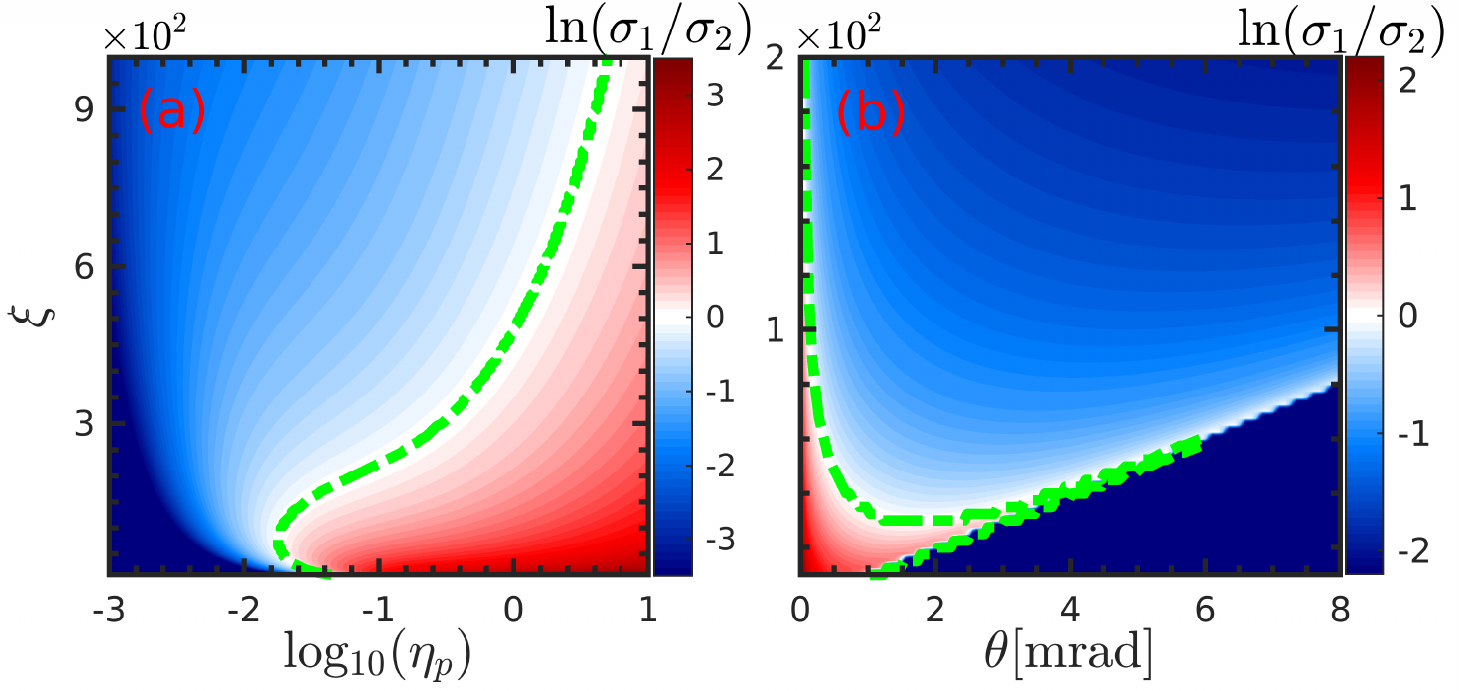}
\caption{Ratio between $\sigma_1$ and $\sigma_2$. (a) $\theta_p=\theta_q=0$, $E_p=E_q$. (b) $\theta_p=-\theta_q=\theta$, $E_p=E_q=10^{4}m$, $\omega_l=4.65~\textrm{eV}$. The green dotted lines denote $\sigma_1=\sigma_2$. $N_l=1$, and the other parameters are same in Fig.~\ref{Fig_wave_packet_inte_F1}.}
\label{Fig_One_Two_photons}
\end{figure}

In Fig.~\ref{Fig_One_Two_photons}, we compare the ratio $\sigma_1/\sigma_2$ of the cross sections for the two processes. As we can see, with small incident angle $\theta\ll 1$, the laser assisted one-photon pair-annihilation can be more probable than the two-photon annihilation in vacuum, especially when we have head-on collision $\theta=0$ with the laser pulse. To measure this in experiment, we see that one would have to resolve the angular spectra of annihilation photons, where, in the small-angle region, one-photon annihilation from within the pulse can exceed two-photon zero-field annihilation.

\section{Numerical Implementation}\label{simulations}
%
In this section, we combine our analytical calculations with numerical implementation. To consider the number of one-photon pair-annihilation events in realistic situations, we specify the positron momentum distribution to be $\left|\rho(q)\right|^2=(2\pi)^3(q^{0}/m)\delta^{(3)}(\bm{q}-\bm{q}_{i})$ which clearly fulfills the normalisation condition Eq.~(\ref{Eq_wave_packet_normalisation}). The number $N_{a}$ of positron annihilation events in the interaction of $N_{e^{+}}$ incident positrons with a dense electron target and a laser pulse is then:
\begin{align}
N_{a}=N_{e^{+}}n_{e^-}\lambda^3_c\frac{\alpha}{2\pi }\frac{m}{p^{0}\eta_q}\int d\psi~ f(p,q,\bm{\xi})\,.
\label{Eq_Annihilation_number}
\end{align}
This number of events is suppressed by the electron density factor, which is small unless there is, on average, one electron per Compton wavelength cubed. (This would correspond to a density of $\sim 7\times 10^{28}~\textrm{cm}^{-3}$, more than $10^4$ times higher than solid density $\sim 10^{24}~\textrm{cm}^{-3}$.) In the following, we consider two example applications of one-photon pair-annihilation.

\subsection{QED Cascade and Laser Plasma Interaction}
\begin{figure}[t!]
\includegraphics[width=0.46\textwidth]{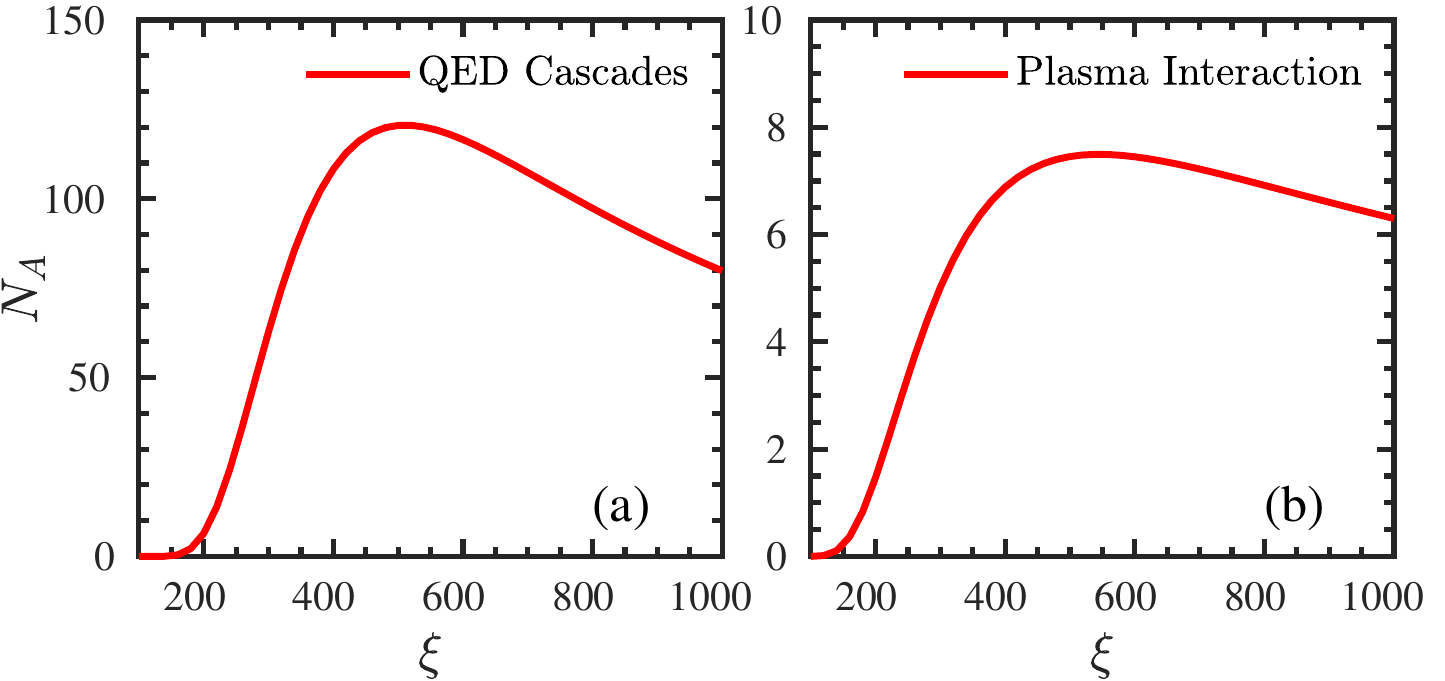}
\caption{ Number of annihilation events in (a) QED cascades, (b) a laser-plasma interaction. A linearly polarised laser pulse: $\bm{\xi}(\psi)=\xi\,\textrm{sech}^2[\psi/(\omega_lT)]\,[\sin(\psi),0,0]$ is employed where $T=5T_l$, $\omega_l=1.55~\textrm{eV}$, $T_l=2\pi/\omega_l$.}
\label{Fig_Cascade_plasma}
\end{figure}

We first consider one-photon pair-annihilation in QED cascades. In Ref.~\cite{PRL2011Nerush} the typical particle density in a QED cascade was given as approximately equal to the relativistic critical density $n_{e^{+}}=n_{e^{-}}\approx \xi n_{c}$, in which $n_c=\omega^2_lm/{4\pi}$ is the plasma critical density. The typical particle energy in the cascade is around $E_{p}\approx E_{q}\approx m\xi$. Given these parameters, we show in Fig.~\ref{Fig_Cascade_plasma} (a) the number of pair annihilations in the volume of one laser wavelength cubed. In the calculation, the number of positrons is $N_{e^{+}}\approx \xi n_{c}\lambda^{3}_l$. As we can see, the number of annihilations is at best six orders of magnitude smaller than the initial positron number. We thus conclude that one-photon annihilation will have a negligible effect on QED cascades.

Another scenario in which a high electron density can arise is the irradiation of a solid plasma with an intense laser pulse~\cite{Pop2009Schlegel,Tang_2019}. At the plasma surface, an extreme density electron foil, with the typical density $n_{e^{-}}\sim \xi^2 n_c$ and energy $E_{p}\approx \xi m$, can be compressed. We consider the number of annihilations when a beam of $N_{e^+}\approx 10^{8}$ positrons with $E_{q}=2000m$ is fired at the electron foil. Fig.~\ref{Fig_Cascade_plasma} (b) shows calculation results for the number of annihilation events during this laser-plasma interaction. Again, this number is many orders of magnitude lower than  the initial number of positrons, and as for cascades we conclude that one-photon pair-annihilation is negligible.

We note that our calculations neglect the influence of the particle direction and assume all the particles move head-on with the laser pulse, in order to consider the most optimistic situation for one-photon pair-annihilation. When more experimentally-realisable parameters are considered, the number of one-photon pair-annihilation events could be much smaller than the estimated numbers.
%
\subsection{Incorporation in PIC}
%
The $1 \rightarrow 2$ quantum processes of NLC and NBW are now commonly included within the contemporary particle-in-cell (PIC) framework~\cite{PRE2015Gonoskov,Arber_2015}, while the $2 \rightarrow 1$ process of one-photon pair-annihilation (and one-photon absorption~\cite{ilderton2019absorption}) is neglected. To incorporate annihilation into the standard PIC-algorithm, we use the so-called probability ``rate'', c.f.~Eq.'s~(\ref{Eq_LCFA_wave_packet})--({\ref{Eq_LCFA_Integrand_F}):
\begin{align}
\frac{d\tsf{P}}{dt}= \alpha\lambda^2_c n_{e^-}\frac{m^2}{\pi^{0}_p \pi^{0}_q} f(p,q,\bm{\xi})\,,
\label{Eq_probability_rate}
\end{align}
which can be implemented in the standard PIC-algorithm as it depends only on local parameters. In each time step $\Delta t$, the probability for one positron annihilated in the $j$th pseudo-positron is
\begin{align}
\tsf{P}_{j}=w_j\Delta t\sum_{i}\frac{w_i}{\Delta V}\frac{\alpha\lambda^2_cm^2}{\pi^{0}_{p_i} \pi^{0}_{q_{j}}} f(p_{i},q_{j},\bm{\xi})=\sum_{i}\tsf{P}_{i,j}\,,
\label{Eq_probability_rate_PIC}
\end{align}
where we sum over all the electrons $i$ in the same grid cell as the $j$th pseudo-positron, $\Delta V$ is the volume of the cell, and $w_{i,j}$ are the particle weights.
A Monte Carlo method is applied to describe the one-photon pair-annihilation process semi-classically. Two random numbers $r_1$ and $r_2$ in $[0,1]$ are generated to determine whether an annihilation occurs and to choose the momentum of the photon. For each pseudo-positron, $j$, an annihilation event is accepted if $r_1<\tsf{P}_{j}$, and then the momentum of the produced photon is calculated using the momentum of the $k^\text{th}$ pseudo-electron, if ${\sum_{i<k-1}\tsf{P}_{i,j}/\tsf{P}_j<r_2\leq\sum_{i<k}\tsf{P}_{i,j}/\tsf{P}_j}$. Because the probability is extremely small, we can ignore the decrease of the particle weight induced by one-photon pair-annihilation.
\begin{figure}[t!]
\includegraphics[width=0.35\textwidth]{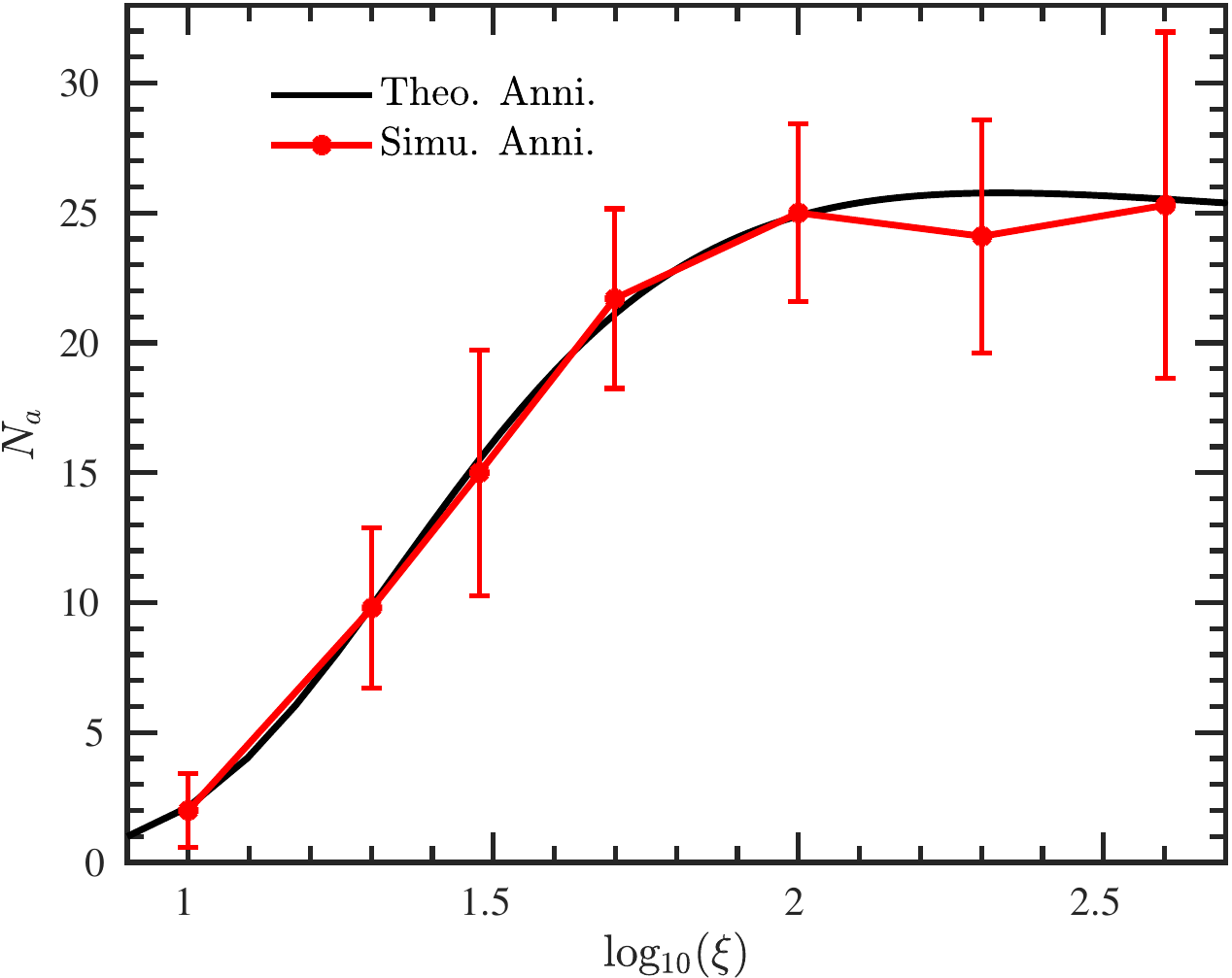}
\caption{Numerical simulations. Black line: theoretical number of annihilations. Red line with error bars: simulation results for the number of annihilations, averaged over $40$ runs, with the error bar denoting the standard deviation. The laser pulse is same as in Fig.~\ref{Fig_Cascade_plasma} except $\omega_l=4.65~\textrm{eV}$.}
\label{Fig_code_number}
\end{figure}

To demonstrate, we implement our method in a single particle code, i.e. we neglect possible plasma effects~\cite{PRSTAB2011Elkina}.
The particle beams have the initial conditions: $N_{e^+}=10^{11}$, $n_{e^-}=10^{25}~\textrm{cm}^{-3}$, $E_p=E_q=2000m$, $\theta_p=\theta_q=0$, and the beam length $0.01\lambda_l$, and the size of grid cell is $\Delta V=\Delta z \pi R^2$ where $\Delta z=2\times 10^{-4}\lambda_l$ and the transverse width of the laser beam is $R=5\lambda_l$.
As shown in Fig.~\ref{Fig_code_number}, the simulated number of annihilations (red line) matches well with the theoretical result (black line).

In our numerical examples, in order to obtain an appreciable number of one-photon pair-annihilations, we have considered a large number of positrons and an extremely dense electron beam. To obtain a prediction for a smaller number of initial positrons and a more realistic less-dense electron beam, the number of one-photon pair-annihilation events can be scaled from the simulation result based on the ratio of positron number and electron density:
\bea
N_{a}^{\prime}  = N_{a}\frac{N_{e^{+}}^{\prime}n_{e^{-}}^{\prime}}{N_{e^{+}}n_{e^{-}}}\,N_{\tsf{sim.}}\,, \label{eqn:NA2}
\eea
where $N_{\tsf{sim.}}$ is the number of simulations. To prove this, we decrease the positron number to $N_{e^{+}}^{\prime}=10^{9}$ and the electron density to $n_{e^{-}}^{\prime}=10^{23}~\textrm{cm}^{-3}$, and repeat the simulations $N_{\tsf{sim.}}=2\times 10^{5}$. We observe $N^{\prime}_a=493$ one-photon pair-annihilation for $\xi=100$, which matches the predicted number $497.6$ from the annihilation number $N_a=24.88$ at $\xi=100$ in the black line in Fig.~\ref{Fig_code_number}.

This method can also be simply extended to realistic situations with specific momentum distributions because of the way the momentum part of the wavepacket factorises into the total expression£¬ see Eq.~(\ref{Eq_LCFA_wave_packet}).
If we consider a simulation with the positron momentum distribution $|\rho(q)|^2$ , we can, in principle, split it into a set of simulations with different positron momenta $q$ and number $N_{e^{+}}|\rho(q)|^2$, and then sum the results $N_a$ in each case.
Based on this point, even though rapid oscillations appear in the exact result in \figref{Fig_LCFA_benchmark}, it is reasonable to implement the LCFA result in a standard PIC-code, because the LCFA effectively averages across these oscillations when implemented in this way.

\section{Conclusion}\label{conclusion}
%
We have analysed one-photon electron-positron pair annihilation in a plane wave background. We derived the locally constant field approximation (LCFA) for this process and benchmarked it against the exact result for a circularly polarised monochromatic background. As one may expect on the basis of LCFA results for NLC, the LCFA was found to be incapable of reproducing harmonic structure. However, a new shortcoming of the LCFA was identified: the LCFA result cannot reproduce the physics of narrow wavepackets, which here manifested as a highly oscillatory structure in the high-energy region.

We obtained simple scaling relations for annihilation in various setups, and compared the one-photon annihilation cross section in a plane wave with the cross section of two-photon pair-annihilation in vacuum. The one-photon process can be dominant for small-angle scattering in the head-on configuration.}

Using numerical simulations based on the LCFA we were able to confirm that one-photon pair-annihilation will have a negligible effect on QED cascades and certain laser-plasma interactions at realisable particle densities. We also showed that annihilation can be included in large-scale numerical simulation frameworks, benchmarking our results against a Particle-In-Cell (PIC) simulation.

\section{Acknowledgments}
We thank A. J. Macleod for useful discussions. The authors are supported by the EPSRC, Grant No.~EP/S010319/1.


\end{document}